\documentclass[12pt]{iopart}

\usepackage{slashed}
\usepackage{braket}
\usepackage{hyperref}

\usepackage{iopams}  
\begin{document}

\title{Hierarchy and decoupling}

\author{Mat\v{e}j Hudec and Michal Malinsk\'{y}}

\address{Institute of Particle and Nuclear Physics,
Faculty of Mathematics and Physics,
Charles University in Prague, V Hole\v{s}ovi\v{c}k\'ach 2,
180 00 Praha 8, Czech Republic}
\ead{hudec@ipnp.troja.mff.cuni.cz, malinsky@ipnp.troja.mff.cuni.cz}
\vspace{10pt}

\begin{abstract}
A large hierarchy between the electroweak scale and virtually any new scale of beyond-Standard-Model physics is often claimed to be unnatural. 
Sometimes, the apparent disparity between the measured Higgs mass and the size of the typical loop corrections encountered within such schemes is even interpretted as a profound indication that one should expect a remedy just behind the corner, probably in form of a new physics such as low-scale supersymmetry, new strong dynamics etc.

In reality, all such potentially large corrections in the one- and two-point Green's functions cancel with each other in the physical Higgs mass $m_{H}$ which eventually turns out to be driven only by the electroweak VEV $v$. This, naively, may look like a miracle, the more that the standard perturbative approach often makes it irresistible to discuss the magnitudes of those corrections as if, individually, they were physically relevant.

To shed some more light on this conundrum we advocate a method based on the symmetry properties of the Coleman-Weinberg effective potential which not only makes it very clear why $m_H\propto v$ to all orders in the perturbative expansion but, at the same time, it does not require any inspection of the explicit form of the tadpole equations whatsoever. Besides simplifying the calculations considerably it makes the ``internal composition'' of the VEV in terms of the high-scale parameters essentially irrelevant.  
 
We  exemplify these findings on an extended series of specific simplified models in which the role of the heavy dynamics is played by all ``reasonable'' types of fields (barring gravity), i.e., by a heavy scalar, a heavy (Majorana) fermion and a heavy vector, respectively.  We show that the dependence of  $m_{H}$ on the heavy scale follows the expectation based on dynamical arguments such as the decoupling theorem. 
\end{abstract}

\section{Introduction}
The Higgs boson discovery in summer 2012 was a true climax of the RUN-I LHC activities and, as such, it brought high expectations for further discoveries in subsequent stages. Unfortunately, there have been no real\footnote{This, however, does not mean at all that RUN II was boring -- the 750 GeV fluke in the di-photon channel sparked an unprecedented rush in 2015-2016 until it eventually faded away; the flavour anomalies in the $B$-meson decays are still around as of now and there is a lot of activity in attempts to understand these peculiarities in terms of, e.g., new light degrees of freedom.   
} 
break-throughs achieved at the front of the  new physics searches in the LHC RUN-II. 
Especially, the lack of any hints of the low-scale supersymmetry (SUSY) is, by many, viewed as particularly disturbing as it was expected to provide answers to many fundamental questions about the apparent shortcomings of the Standard Model (SM) such as the nature and origin of the dark matter or the size of the baryonic fraction of the critical density of the Universe. 

On the theory side, perhaps the most prominent issue the low-energy SUSY was expected to address was the so called hierarchy problem, see, e.g.,~\cite{Giudice:2008bi} and references therein. It is usually phrased as a question of why, in the stipulated presence of new physics at very high energies (such as the GUT or the Planck scale), the electroweak scale is so incomparably smaller.

Concerning multi-scale QFT's in general, it is very traditional to parametrize these frameworks by their heavy-sector parameters because it is believed that physical theories should always ``explain'' \emph{macroscopic} phenomena in terms of the underlying \emph{microscopic} laws. The hierarchy problem then becomes an issue of \emph{fine-tuning} since a few large parameters  must typically combine into much smaller quantities.
As an illustration, consider the minimal $SU(5)$ GUT~\cite{Georgi:1974sy} where the hierarchy problem emerges as the issue of the doublet-triplet splitting. Here, the mass of the SM-like Higgs boson is given by \begin{equation}\label{large}
m^2_H \propto \mu_5^2 + \eta V^2
\end{equation}
where $\mu_{5}^{2}$ is the quadratic self-coupling of the fundamental scalar 5-plet, $V\sim 10^{16}$ GeV is the large VEV of the 24-dimensional adjoint scalar and $\eta$ is the $5-24$ scalar potential quartic coupling. 
Obviously, the LHS of Eq.~(\ref{large}) with $m_H\approx 125$ GeV is unstable with respect to small relative changes of the input parameters  $\mu_5$, $\eta$ and $V$, which technically means that the (dimensionless) logarithmic derivatives~\cite{Barbieri:1987fn}
\begin{equation}
\frac{\partial \log m^2_H}{\partial \log \mu_{5}^2} = \frac{m_{H}^{2}-\eta V^2}{m^2_H} 
\quad {\rm and}\quad 
\frac{\partial \log m^2_H}{\partial \log \eta}= \frac{\partial \log m^2_H}{\partial \log V^2}=\frac{\eta V^2}{m^2_H}
\end{equation}
turn out to be large when the real-world data are inserted.

Remarkably enough, things get much better (at least optically) if the ambition of parametrizing everything in terms of just the high-energy inputs is revoked. Indeed, once we use the electroweak VEV $v$ instead of $\mu_{5}$, Eq.~(\ref{large}) is simplified (in accordance with the minimal survival hypothesis~\cite{delAguila:1980at,Mohapatra:1982aq}) into 
\begin{equation}
m^2_H \propto \lambda v^{2}\,,
\end{equation}
where $\lambda$ is the quartic self-coupling of the scalar 5-plet, 
and no instabilities in predictions of other observables arise even if $v \ll V$:
\begin{equation}\label{small}
\frac{\partial \log m^2_H}{\partial \log V^2} =0
\quad {\rm and}\quad 
 \frac{\partial \log m^2_H}{\partial \log v^2}=1. 
\end{equation}
This elementary twist is, however, often 
dismissed as a mere rhetoric exercise, the more that the central technical question of interest, namely,
``What makes $v^{2}$ which, up to an ${\cal O}(1)$ coupling, is formally a sum of two large terms in (\ref{large}) so small?'' remains unanswered. 

In writing~(\ref{small}) we have implicitly touched upon another aspect of the usual misunderstanding related to the hierarchy issue which corresponds to the 
claim that even with $v$ fixed at  the electroweak scale the mass of the Higgs  \emph{particle} should be lifted to the higher scale $\Lambda \gg v$ due to the large quantum corrections to the Higgs field two-point function formally proportional to $\Lambda^2$. 

This, however, does not look correct as in such a case, one would immediately run into a flagrant conflict with the decoupling theorem~\cite{AppelquistCarazzone} which states that in the limit $\Lambda\to\infty$ the SM as a renormalizable theory should become insensitive to any high-energy sector (assuming its couplings with the light one are well behaved). In other words, there should be only one scale to which the Higgs mass may still be proportional, namely, the low-scale VEV. 

This apparent paradox has been partially addressed in, e.g., Ref.~\cite{fun} where the situation has been thoroughly analyzed in the  Abelian Higgs model at the one-loop effective potential level. In particular, it was shown that besides the two-point function $\Gamma^{(2)}$, the $\Lambda^2$-proportional corrections influence also the tadpole equation $\Gamma^{(1)}=0$; this, in turn leads to a ``miraculous'' cancellation among these corrections in the formula for the physical Higgs mass.

In what follows, we shall extend the previous analysis~\cite{fun} to other cases of interest where the large-scale dynamics is realized in several different forms (including two popular extensions of the SM).
In all these settings, the explicit proportionality of the mass of the lightest neutral scalar to the corresponding VEV is, indeed, revealed. 
In doing so, we employ a method based on the symmetry properties of the Coleman-Weinberg effective potential which goes beyond that advocated in~\cite{fun}. It not only makes it especially clear why $m_H\propto v$ but, on top of that, it does not require any inspection of the explicit form of the tadpole equations and, thus, makes the ``internal composition'' of the electroweak VEV in terms of the high-scale parameters essentially irrelevant.

 \vskip 2mm

The manuscript is organised as follows: In Sections~\ref{sect:prerequisites} and~\ref{sec:AbelianModel} we explain the merits of the ``gauge invariant'' approach advocated in this work along with the necessary definitions of the basic structures and a simple example of the EP method applied to the masses of the light spin-0 field(s) in the scalar electrodynamics in both the symmetric as well as the Higgs phase. Subsequently, in Sections~\ref{sec:superconductor}--\ref{H&Fermion} we provide a set of further examples which extend the limited scope of the previous work~\cite{fun} to essentially all remaining potentially interesting cases, i.e., to the situations where the heavy sector is represented by a massive vector boson and/or a (Dirac or Majorana) heavy fermion. Section~\ref{sec:HTM} is devoted to a more general case of the triplet Higgs model (aka type-II seesaw) where the simplifying arguments based on the generalised custodial symmetry that were heavily used in preceeding parts can not be used in a straightforward manner; nevertheless, the core of the argument remains intact. 
\section{Prerequisites}\label{sect:prerequisites}
\subsection{The effective potential approach}
The quantum corrections to the scalar spectrum of a given model are conveniently calculated by the effective potential techniques. The central object of interest is then the matrix of second derivatives of the effective potential $V_\mathrm{eff}(\varphi)$ evaluated at its minimum. Sticking, for the moment, to the single-field case one has
\begin{equation}
{\mathfrak{m}^2} = \left\langle\frac{\partial^2 }{\partial\varphi^{2}}V_\mathrm{eff}\right\rangle\,,
\label{eq:MassMatrixDefinition}
\end{equation}
where the angle brackets denote the vacuum field configuration conforming the stationarity condition  
\begin{equation}
\left\langle \frac{\partial V_\mathrm{eff}}{\partial \varphi}\right\rangle = 0.
\label{eq:StationarityCondition}
\end{equation}
We shall mostly ignore the fact that $\mathfrak{m}^2$ is, strictly speaking, not the physical mass of the relevant scalar as it corresponds to the value of the relevant 1PI Green's function (calculated in the $\overline{\rm MS}$ scheme) evaluated at zero external momentum rather than its true $p^{2}$-root. For what follows this nuance is, indeed, irrelevant as the terms one needs for the transition to the pole masses are generally harmless. A more detailed account of this point can be found in the literature; see, e.g., \cite{Quiros}.

\subsection{Non-linear coordinates}\label{sec:gaugecoords}
In some cases, prescription (\ref{eq:MassMatrixDefinition}) may not provide the most convenient approach to the mass  calculation. Indeed, if the effective potential happens to depend solely on the square\footnote{Needless to say the 'square' here may be easily generalized to structures like $\varphi^{\dagger}\varphi$ often encountered in the SM and its simple extensions.} of a certain field $\varphi$  then its first two derivatives with respect to this coordinate may be written as 
\begin{equation}
\frac{\partial}{\partial \varphi} V_\mathrm{eff}(\varphi^2) = 2\varphi \, \frac{\partial V_\mathrm{eff}}{\partial (\varphi^2)}\,,
\label{eq:MasterEquation1}
\end{equation}
\begin{equation}
\frac{\partial^2}{(\partial \varphi)^2} V_\mathrm{eff}(\varphi^2) = 2\, \frac{\partial V_\mathrm{eff}}{\partial (\varphi^2)} + 4 \varphi^2 \frac{\partial^2V_\mathrm{eff}}{\left[\partial(\varphi^2)\right]^2}\,.
\label{eq:MasterEquation2}
\end{equation}
Note that this system of equations behaves in a rather different way if one looks for its solutions corresponding to the symmetric and asymmetric minima, respectively.
As for the former (symmetric) case, Eq.~(\ref{eq:MasterEquation1}) just says there is always a local extremum at $\varphi=0$ and the mass of the associated scalar is given by the first term on the RHS of  Eq.~(\ref{eq:MasterEquation2}). 
On the other hand, in any asymmetric (local) minimum the zero of the LHS of Eq.~(\ref{eq:MasterEquation1}) implies that the first term on the RHS of Eq.~(\ref{eq:MasterEquation2}) drops and the corresponding mass (squared) matrix element is proportional to the (square of the) VEV of $\varphi$ (to be called $v$). This, per se, makes it very clear that {\em the mass of the Higgs boson is always proportional to its VEV}, namely
\begin{equation}
\label{eq:MasterEquation3}
{\mathfrak m}^{2}=4 v^2 \left\langle\frac{\partial^2V_\mathrm{eff}}{\left[\partial(\varphi^2)\right]^2}\right\rangle
\,.
\end{equation}

The only potential loophole to this argument is the possibility that the dimensionless bracket on the RHS of  Eq.~(\ref{eq:MasterEquation3}) may, in principle, factor out a negative power of $v$. This, however, is not the case; the point is that ${\partial^2V_\mathrm{eff}}/{\left[\partial(\varphi^2)\right]^2}$ is tightly connected (see~\ref{App:Amplitudes}) to the physical (zero-momentum) 1PI Green's functions $\Gamma^{(n)}$ ($n\geq 3$), namely 
\begin{equation}
\left\langle\frac{\partial^2V_\mathrm{eff}}{\left[\partial(\varphi^2)\right]^2}\right\rangle=\frac{1}{4}\sum_{n=3}^{\infty} 
\frac{(-1)^{n+1}}{\, (n-1)!}\,
v^{n-4}\,\Gamma^{(n)}(0)\,
,
\label{eq:InvertedAmplitudes}
\end{equation}
whose dependence on the heavy scale $M$ (and, thus, also on $v$) is not arbitrary. 
Indeed, the scaling of the $M$-dependent parts (to be indicated by $\Delta_{M}$) of  $\Gamma^{(n)}(0)$ is determined\footnote{This follows from the fact that there are no other scales at play but $v$ and $M$.} by the requirement of decoupling~\cite{AppelquistCarazzone} to be, at worst,
\begin{equation}
\label{Gamma}
\Delta_{M}\Gamma^{(n)}(0)\propto\frac{v^{4-n+\alpha}}{M^{\alpha}}\, f^{(n)}(v,M)\,,
\end{equation}
where $\alpha\geq0$ and $f^{(n)}(v,M)$ are dimensionless functions which are ``well behaved'', i.e., at most asymptotically flat in the $M\to \infty$ limit.
Hence,
\begin{equation}
\Delta_{M}
\left\langle\frac{\partial^2V_\mathrm{eff}}{\left[\partial(\varphi^2)\right]^2}\right\rangle
\propto \left(\frac{v}{M}\right)^{\alpha}
\end{equation}
and the heavy scale contribution to the (zero-momentum) mass of the associated Higgs boson scales with $v$ at least as
\begin{equation}
\Delta_{M}\mathfrak{m}^{2}\propto v^{2}\left(\frac{v}{M}\right)^{\alpha}\,.
\end{equation}
As a matter of fact, this is nothing but the behaviour expected on the basis of the decoupling theorem~\cite{AppelquistCarazzone} applied directly to the two-point function; it has just been conveniently re-phrased in a language of  scattering amplitudes which, as we believe,  leaves less room for misconception. 

Several remarks are worth making here: First, it is well known that, formally, the decoupling behaviour of low-energy observables is apparent only in the momentum schemes.  In other schemes such as the $\overline{\rm MS}$, the $f^{(n)}(v,M)$ functions in formulas of the type~(\ref{Gamma}) may exhibit a weak (typically logarithmic) rudimentary growth with $M\to \infty$ which, however, just reflects the unphysicality of the  scheme (i.e., the renormalized parameters) at play.

Second, the simple setting that we have been entertaining above may be trivially generalized to more realistic scenarios with more than a single field in the scalar sector. In such a case,
the second derivatives populate a matrix $\mathcal{M}^2$
and the renormalized ($\overline{\rm MS}$) scalar masses then correspond to its eigenvalues (to be denoted by $\mathfrak{m}^{2}_i$).  Similarly, the $\varphi^{2}$ structure would then typically (though not necessarily) correspond to quadratic scalar gauge invariants.

\subsection{Further comments}
In what follows, we shall demonstrate these arguments on a set of examples based on sample models of spontaneous symmetry breaking (SSB) in which the role of the heavy sector will be played by various types of fields. For the sake of simplicity we shall stick to the one-loop approximation to $V_{\rm eff}$ \`a la Coleman and Weinberg \cite{ColemanWeinberg}. 
This, in the Landau gauge (and in the $\overline{\rm MS}$ scheme) reads 
$V_\mathrm{eff} = V_0+V_{\rm S}+V_{\rm F}+V_{\rm G}$ where $V_{0}$ is the classical potential and
\numparts
\label{eqs:VSFG}
\begin{eqnarray}
V_\mathrm{S} &= \;\frac{1}{64 \pi^2} \; \mathrm{Tr}\!\left[ M_\mathrm{S}^4 \left( \log \frac{M_\mathrm{S}^2}{\mu^2} - \frac{3}{2} \right) \right], \label{VS} \\
V_\mathrm{F} &=\! -\frac{2}{64\pi^2} \mathrm{Tr}\!\left[ (M_\mathrm{F} M_\mathrm{F}^\dagger)^2 \left( \log \frac{M_\mathrm{F} M_\mathrm{F}^\dagger}{\mu^2} - \frac{3}{2} \right) \right]\,,
\label{VF}\\
V_\mathrm{G} &= \; \frac{3}{64 \pi^2} \; \mathrm{Tr}\!\left[ M_\mathrm{G}^4 \left( \log \frac{M_\mathrm{G}^2}{\mu^2} - \frac{5}{6} \right) \right] \label{VG} 
\end{eqnarray}
\endnumparts
denote the contributions from scalars, two--component fermions and gauge fields running in the loops, respectively. Here, $M_{\mathrm{S,G}}^2$ and $M_\mathrm{F}$ are the corresponding scalar-field-dependent mass matrices and $\mu$ is the renormalization scale. Let us also note that the traces above are taken only over the flavour indices.

The models that we shall be dealing with below are defined by their classical action. On the other hand, the arguments in Sect.~\ref{sec:gaugecoords} rely on the symmetry properties of the entire effective potential which do not need to be the same as those of $V_{0}$.
To this end, two remarks are worth making here: First, the global symmetries of the classical action, including the spacetime--independent versions of gauge symmetries, are respected by the Landau--gauge--fixing term
\begin{equation}
\fl\qquad
\mathcal{L}_\mathrm{g.f.}^\mathrm{Landau}=\lim_{\xi\rightarrow 0} \frac{1}{2\xi}\left(\partial \cdot A^a\right)\left(\partial \cdot A^a\right)\,,
\end{equation}
and, hence, the quantum effective potential \`a la Coleman and Weinberg~\cite{ColemanWeinberg} is invariant under all internal\footnote{I.e., those that mix the fields among themselves.} symmetries of the classical action\footnote{Do not confuse gauge symmetry (of the potential w.r. to the gauge transformation) with a gauge dependence (i.e., the dependence on the choice of the parameter $\xi$).}.

Second, though the classical potential may have a larger symmetry than the entire classical action, this feature does not \emph{automatically} transfer to the loop level, i.e., to the effective potential and the effective action. 
Nevertheless, in all cases that we shall be considering in the following sections we shall see that all symmetries of the classical potential are retained even at the quantum level. This, in turn, will simplify our calculations considerably.

\section{The Abelian Higgs model with an extra heavy singlet scalar}
\label{sec:AbelianModel}
As a first example we shall study the electrodynamics of a complex scalar field $\mathcal{X}=(\chi_1 + i \chi_2)/\sqrt{2}$ which interacts with a heavy neutral CP--odd scalar~$\phi$ via the usual portal coupling~\cite{Patt:2006fw}, and study the influence of this singlet on the mass of $\mathcal{X}$ in both symmetric and broken phases.
The tree-level scalar potential of this system reads
\begin{equation}
V_0 = 
\frac{1}{2} m_{\chi }^2\, \chi ^2
+\frac{1}{4}\rho \, \chi ^4
+ \frac{1}{2} m_{\phi }^2 \,\phi ^2 
+\frac{1}{4}\lambda \, \phi ^4
+\frac{1}{2} \kappa \, \chi ^2 \phi ^2\,,
\label{V0}
\end{equation}
where \begin{equation}
\chi^2 \equiv \chi_1^2 + \chi_2^2.
\label{eq:ChiQuadraticInvariant}
\end{equation} The one--loop effective potential in this model can be easily computed by hand, see~\ref{App:U1Veff}. The crucial thing is that, as the effective potential in the Landau gauge inherits the 
$O(2)_\chi \otimes {\mathbb{Z}_{2}}_{\phi}$ symmetry of the classical action, it can be conveniently written as
\begin{equation}
V_\mathrm{eff}=V_\mathrm{eff}(\chi^2, \phi^2)
\label{eq:AHMphiIsSymmetric}
\end{equation}
because there are no other algebraically independent invariants.

\subsection{The unbroken phase}
In the unbroken phase, i.e., for
$\langle\phi\rangle = \langle\chi^2\rangle=0$,
the mass matrix in the $\{\phi,\chi_1,\chi_2\}$ basis takes a diagonal form
\begin{equation}
\mathcal{M}^{2} = 
\left(\!\begin{array}{ccc}
2 \left\langle\frac{\partial V_\mathrm{eff}}{\partial\left(\phi^2\right)}\right\rangle &0&0\\
0&2 \left\langle\frac{\partial V_\mathrm{eff}}{\partial\left(\chi^2\right)}\right\rangle &0\\
0&0&2 \left\langle\frac{\partial V_\mathrm{eff}}{\partial\left(\chi^2\right)}\right\rangle 
\end{array}\right).
\label{eq:AHM:MassMatrixunbroken}
\end{equation}
Note that its particular shape follows from a trivial multi-field generalization of Eqs.~(\ref{eq:MasterEquation1}) and~(\ref{eq:MasterEquation2}) and the discussion underneath.

Using the explicit form of $V_\mathrm{eff}$ given in (\ref{eq:U1Veff}) the one--loop masses of the neutral and charged scalars read
\numparts
\label{ZeroExample:Masses}
\begin{eqnarray}
\fl\qquad
\mathfrak{m}^2_\phi = m_\phi^2 + \frac{3 \lambda}{16\pi^2}m_\phi^2
 \left(\log\!\left[\frac{m_\phi^2}{\mu^2}\right]-1\right)+\frac{\kappa}{8\pi^2} m_\chi^2 \left(\log\!\left[\frac{m_\chi^2}{\mu^2}\right]-1\right)
,
 \\
\fl\qquad
\mathfrak{m}^2_\chi= m_\chi^2 
+
\frac{ \rho}{4 \pi^2} m_\chi^2 \,\left(\log\!\left[\frac{m_\chi^2}{\mu^2}\right]-1\right)
+ \frac{\kappa}{16\pi^2} m_\phi^2 \left(\log\!\left[\frac{m_\phi^2}{\mu^2}\right]-1\right).
\end{eqnarray}
\endnumparts
Notice that both physical masses grow linearly with both mass parameters $m^2_{\phi,\chi}$ and thus, if $\mathfrak{m}^2_\chi \ll \mathfrak{m}^2_\phi$ is required,
the mass parameters must be fine--tuned; from this perspective a hierarchical spectrum is technically unnatural.

\subsection{The Higgs phase}
The scenario where the gauge symmetry is spontaneously broken by the VEV of $\mathcal{X}$ has already been investigated in great detail in~\cite{fun}. We shortly reiterate here the salient points in order to demonstrate the practical advantage of the "invariant approach" advocated in Sect.~\ref{sec:gaugecoords}.

Without loss of generality, one can ascribe the VEV of $\mathcal{X}$ to its first component, i.e. 
\begin{equation}
\langle\phi\rangle=0,\qquad\langle \chi_1 \rangle=v_\chi, \qquad  \langle \chi_2 \rangle=0,
\label{eq:AHM:VEVy}
\end{equation}
which, due to (\ref{eq:AHMphiIsSymmetric}), yields the following shape of the quantum-level mass matrix:
\begin{equation}
\mathcal{M}^{2} = 
\left(\!\begin{array}{ccc}
2 \left\langle\frac{\partial V_\mathrm{eff}}{\partial\left(\phi^2\right)}\right\rangle &0&0\\
0&4v_\chi^2\left\langle \frac{\partial^2V_\mathrm{eff}}{[\partial\left(\chi^2\right)]^2}\right\rangle&0\\
0&0&0
\end{array}\right).
\label{eq:AHM:MassMatrixbroken}
\end{equation}
As expected on the basis of the general discussion in Sect.~\ref{sec:gaugecoords} the mass of the Higgs particle corresponding to the shifted field $h=\chi_1-v_\chi$ is proportional to $v_\chi$. One eventually obtains (cf.~\ref{App:U1Veff})
\begin{eqnarray} 
\fl \qquad
\mathfrak{m}_{h}^2=\mathcal{M}^2_{\chi_1 \chi_1} = 
2\rho v_\chi^2 
+ \frac{ v_\chi^2}{8\pi^2} \Bigg( &
 \rho ^2 \log\!\left[\frac{\rho  v_\chi^2+m_\chi^2}{\mu ^2}\right]
+9 \rho ^2 \log\!\left[\frac{3 \rho  v_\chi^2+m_\chi^2}{\mu ^2}\right]\nonumber \\&
+\kappa ^2 \log\!\left[\frac{m_\phi^2+\kappa  v_\chi^2}{\mu ^2}\right]+2e^4 + 3 e^4\log\!\left[\frac{e^2 v_\chi^2}{\mu^2}\right]
\Bigg),
\end{eqnarray}
in agreement with the corresponding expression in \cite{fun}.

\section{The Higgs boson and a massive vector}
\label{sec:superconductor}
In Section \ref{sec:AbelianModel}, the large-scale physics was represented by a single heavy scalar $\phi$. 
As a next (toy) model, we consider again the scalar electrodynamics broken by the VEV of $\mathcal{X}$ as in Sect.~\ref{sec:AbelianModel}; however, in  the current case we want the large-scale sector of the scheme to be mimicked by a heavy vector field. This, in turn, requires an extended gauge symmetry which gets spontaneously broken by an extra scalar $\Phi$ at the scale corresponding to its VEV $v_{\phi}$. 

Barring the $U(1)^{2}$ option with its notorious gauge-mixing issues~\cite{delAguila:1988jz} the  simplest such scenario corresponds to a SM-like high-scale $SU(2)\otimes U(1)_{Y}$ symmetry broken in two steps like
\begin{equation}
SU(2)\otimes U(1)_Y \stackrel{v_{\phi}}\longrightarrow U(1)_Q  \stackrel{v_{\chi}}\longrightarrow \{1\}\,.
\end{equation}
In such a case the $\Phi$ field may be taken as an $SU(2)$ doublet
\begin{equation}
\label{Phistructure}
\Phi = \frac{1}{\sqrt{2}}\left(\!\begin{array}{cc}
\phi_1 + i\phi_2 \\ 
\phi_3 + i \phi_4 \end{array} \!\right)\,;
\end{equation}
the corresponding $U(1)_{Y}$ (hyper-)charges are chosen as $Y_\phi=\frac{1}{2}$ and $Y_\chi=2$, respectively\footnote{It is perhaps interesting to note that the setting under consideration may be, physically, viewed as a simple model of the SM Higgs effect at $v_{\phi}$ followed by a superconductor phase developed at a very low scale~$v_{\chi}$.}.

The scalar spectrum of the broken-phase theory should, in turn, consist of two massive fields: a heavy SM-like $H$ with the mass of the order of $v_\phi$ and a \emph{light} Higgs $h$ with a mass $\mathfrak{m}_{h}\sim v_\chi$, as expected on the basis of the minimal survival hypothesis~\cite{delAguila:1980at,Mohapatra:1982aq}. The remaining 4 scalar degrees of freedom become the longitudinal components of $W^\pm, Z$ and $\gamma$ where the last acquires its mass after the SSB of $U(1)_Q$ only.

The main point of what follows is to argue that this qualitative picture, in particular, the large hierarchy between the masses of the two different Higgs fields $H$ and $h$, remains intact even when the loop corrections are included.
Even with a more complicated theory than that in Sect.~\ref{sec:AbelianModel} the most general renormalizable scalar potential here still retains the form of~(\ref{V0}) provided
\begin{equation}
\phi^2 = \phi_1^2 + \phi_2^2 + \phi_3^2 + \phi_4^2\,,\qquad \chi^2=\chi_1^2+\chi_2^2.
\end{equation}
Note that the apparent $O(4)_\phi\otimes O(2)_\chi$ symmetry of the corresponding $V_{\rm eff}$ may be viewed as a~\emph{generalized custodial symmetry}. Again, since $\phi^2$ and $\chi^2$ are the only two algebraically independent invariants built purely from the scalar fields at hand, such a symmetry is preserved in the Landau gauge even at the quantum level.
Without loss of generality, one can again ascribe the VEVs arbitrarily to one component of each multiplet, say 
$
\langle\phi_3\rangle = v_\phi$ and $\langle \chi_1\rangle=v_\chi
$.
A simple generalization of the arguments in Sect.~\ref{sec:gaugecoords} then yields the scalar mass matrix in a very suggestive form 
\begin{equation}
\mathcal{M}^2 = \left(\begin{array}{cccccc}
0&0&0&0&0&0\\
0&0&0&0&0&0\\
0&0&4 v_\phi^2\left\langle\frac{\partial^2 V_\mathrm{eff}}{(\partial\phi^2)^2}\right\rangle
&0&4 v_\phi v_\chi \left\langle\frac{\partial^2 V_\mathrm{eff}}{(\partial\phi^2)(\partial\chi^2)}\right\rangle
&0\\
0&0&0&0&0&0\\
0&0&
4 v_\phi v_\chi \left\langle\frac{\partial^2 V_\mathrm{eff}}{(\partial\phi^2)(\partial\chi^2)}\right\rangle
&0&
4 v_\chi^2\left\langle\frac{\partial^2 V_\mathrm{eff}}{(\partial\chi^2)^2}\right\rangle
&0\\
0&0&0&0&0&0
\end{array}\right).
\label{SU2xU1QuantumM2}
\end{equation}
At the tree level, one obtains
\begin{equation}
\label{SU2xU1ddV0}
\left\langle\frac{\partial^2V_0}{(\partial\phi^2)^2}\right\rangle = \frac{\lambda}{2},
\qquad
\left\langle\frac{\partial^2V_0}{\partial\phi^2\,\partial\chi^2}\right\rangle = \frac{\kappa}{2},
\qquad
\left\langle\frac{\partial^2V_0}{(\partial\chi^2)^2}\right\rangle=\frac{\rho}{2}\,;
\end{equation}
the corresponding one loop corrections can be found in \ref{App:Vecors}.
Combining these findings one ends up with 
\numparts
\begin{eqnarray}
\mathfrak{m}^2_{H} &= 4 \,V_{11}\, v_\phi^2 + \Or\!\left(v_\chi^2 \log v_\phi^2\right),
\label{eq:superCond_H_mass}
\\
\mathfrak{m}^2_{h} &= 4\, \frac{V_{11} V_{22}-V_{12}^2}{V_{11}}\, v_\chi^2 + \Or\!\left(\frac{v_\chi^4}{v_\phi^2}\, \log v_\phi^2\right),\label{eq:superCond_h_mass}
\end{eqnarray}
\endnumparts
where an abbreviation
\begin{equation}
V_{11} \equiv \left\langle\frac{\partial^2V_\mathrm{eff}}{(\partial\phi^2)^2}\right\rangle,
\qquad
V_{12} \equiv \left\langle\frac{\partial^2V_\mathrm{eff}}{\partial\phi^2\,\partial\chi^2}\right\rangle,
\qquad
V_{22} \equiv \left\langle\frac{\partial^2V_\mathrm{eff}}{(\partial\chi^2)^2}\right\rangle.
\label{SU2xU1ddVeffDef}
\end{equation}
has been used. 
As expected, the $v_\chi$ dependence of the corrections~(\ref{SU2xU1ddVeffDef}) is such that the \emph{light} Higgs mass $\mathfrak{m}_{h}$ remains proportional to $v_{\chi}$ even when the loop corrections from the heavy gauge bosons and scalars are incorporated.

\section{The SM Higgs boson and a right-handed neutrino}\label{H&Fermion}

Let us now take the game one step furhter and consider the SM-like setting from the previous section\footnote{From now on we shall be interested in the mass of the physical Higgs boson $H$ and, hence, in what follows we shall consider only the first step of the symmetry breaking discussed in Sect.~\ref{sec:superconductor}.}  as a low-energy theory and discuss the fate of the  Higgs boson mass in the context of a pair of its high-energy extensions motivated by the popular seesaw models of type I and II. 

First, let us consider a model in which the heavy sector contains an extra Weyl fermion $\nu^{c}_{L}$ resembling the RH neutrino. The tree-level scalar potential then remains just that of the SM, i.e.,
\begin{equation}
V_{0} = m^2_\phi \Phi^\dagger \Phi + \lambda \left(\Phi^\dagger \Phi\right)^2,
\end{equation}
with $\Phi$ as in (\ref{Phistructure}).
As such, the scalar sector exhibits an $O(4)$ custodial symmetry which implies that $\mathfrak{m}_{H}$ obeys Eq.~(\ref{eq:MasterEquation3}) that reads here
$${\mathfrak m}_{H}^{2}=4 v_{\phi}^2 \left\langle\frac{\partial^2V_\mathrm{eff}}{\left[\partial(\phi^2)\right]^2}\right\rangle
\,.
$$
For simplicity, we shall work with just a single generation of leptons. The relevant Lagrangian then reads
\begin{equation}
\mathcal{L}=\mathcal{L}_\mathrm{SM} + i \overline{\nu^c_{\,L}}
\slashed{\partial}\, \nu^c_{\,L} -\left( y_\nu \,
{\nu^c_{\,L}}^T \mathcal{C}^{-1} L \,\varepsilon \Phi  + \frac{1}{2}  m_R\, {\nu^c_{\,L}}^T \mathcal{C}^{-1} \nu^c_{\,L} + \mathrm{h.c.}\right)\,,
\end{equation}
where $L$ denotes the lepton doublet, $\varepsilon$ stands for the antisymmetric tensor in the $SU(2)_L$ space and $m_R \gg v_\phi$ sets the scale of new physics. 
By virtue of the $O(4)$ symmetry one can again work with a reduced set of degrees of freedom, namely, $\left(\phi_1,\phi_2,\phi_3,\phi_4\right)\rightarrow \left( 0,0,\phi,0\right)$. The neutrino mass matrix -- a submatrix of $M_\mathrm{F}$ in Eq.~(\ref{VF}) -- then obeys
\begin{equation}
M_\nu(\phi) = 
\left(
\begin{array}{cc}
0  &  y_\nu \phi /\sqrt{2}\\
y_\nu \phi /\sqrt{2} & m_R
\end{array}\right)
\end{equation}
and the corresponding neutrino contribution to the effective potential reads
\begin{equation} 
\label{Vnu}
\fl \quad
V_\nu = -\frac{1}{512\pi^2}\sum_{\pm}
\left(m_R \pm \sqrt{m_R^2+2 \phi^2 y_\nu^2}\right)^4
\left(
2\log\,\vline\frac{m_R\pm \sqrt{m_R^2 + 2 \phi^2 y_{\nu }^2}}{2 \mu}\vline-\frac{3}{2}
\right).
\end{equation}
Note that we neglect the loop corrections from the rest of the SM fields as they do not have anything to do with $m_R$ but rather with $v_{\phi}$. 
With this at hand the Higg mass formula (\ref{eq:MasterEquation3}) with $V_\mathrm{eff} \approx V_0+V_\nu$ may be evaluated readily:
\begin{equation}
\mathfrak{m}_{H}^2= 2\lambda v_\phi^2  
-\frac{y_{\nu}^4 v_\phi^2}{8\pi^2}
\left(
1+\log\frac{m_R^2}{\mu^2}
\right)+\Or\!\left(\frac{v_\phi^4}{m_R^2}\right)
\end{equation}
The proportionality of the Higgs mass to the square of the associated VEV has been attained as expected; the weak residual logarithmic growth of  $\mathfrak{m}_{H}$ with $m_R$ is again attributed to the unphysicality of the renormalization scheme at play.

It is perhaps interesting to note that the fermionic loop correction (\ref{Vnu})  is generally large and negative and, as such, it can per se lead to a spontaneous breakdown of the SM gauge symmetry even if $m^2_\phi > 0$.

\section{The Higgs triplet model}\label{sec:HTM}
The effective potentials studied in the preceding sections all obeyed a variant of the \emph{(generalized) custodial symmetry} which turned out to be a very powerful tool for the analysis of the Higgs mass sensitivity to the stipulated new physics at some very large scale.
In the remainder of this study we shall turn our attention to a setting without this symmetry; as we shall see, even in such a case a similar trick can still lead to a considerable simplification of the matters. 

Let us consider the scalar sector of the type-II seesaw model~\cite{PhysRevD.22.2860, PhysRevD.22.2227} which consists of an $SU(2)$ doublet $\Phi \!\sim\! (1,2, +1/2)$ and an $SU(2)$ triplet $\Delta\!\sim\!(1,3,+1)$. It is convenient to represent this system by a two-component vector and a $2\times 2$ traceless matrix, respectively, which accommodate the electric charge  ($Q=T^3_L + Y$) eigenstates as
\begin{equation}
\Phi =\left(\begin{array}{c}\phi^+\\ \phi^0\end{array}\right),\qquad \Delta = \left(\begin{array}{cc}\Delta^+ / \sqrt{2}  & \Delta^{++} \\ \Delta^0 & -\Delta^+/\sqrt{2}
\end{array}\right).
\end{equation}
The most general renormalizable tree-level potential through which these fields can interact reads
\begin{eqnarray}\fl
\label{HTMpot}
V_0 = m_\mathrm{\phi}^2 \Phi^\dagger \Phi
+ M_\Delta^2 \mathrm{Tr}[\Delta^\dagger \Delta ]
+ \left(\nu\,
\Phi^T \varepsilon \Delta^\dagger \Phi + \mathrm{h.c.}\right) 
+ \lambda_1(\Phi^\dagger \Phi)^2 
\\+ \lambda_2\left(\mathrm{Tr}[\Delta^\dagger\Delta] \right)^2 
+ \lambda_3 \mathrm{Tr}[\Delta^\dagger\Delta \Delta^\dagger\Delta]
+\lambda_4 (\Phi^\dagger \Phi)\mathrm{Tr}[\Delta^\dagger\Delta]
+\lambda_5 \Phi^\dagger \Delta \Delta^\dagger \Phi. \nonumber
\end{eqnarray}
Notice that the would--be $O(4)_\Phi \otimes O(6)_\Delta$ generalized custodial symmetry is explicitly violated by the $\nu$, $\lambda_3$ and $\lambda_5$ terms. 
Without loss of generality the VEV of $\phi^{0}$ as well as the $\nu$ parameter can be both made real by field redefinitions.  
Note also that a non-zero $\langle \phi^{0} \rangle\equiv v_\phi/\sqrt{2}$ implies a non-zero $\langle \Delta^{0}\rangle \equiv v_\Delta/\sqrt{2}$; the two are related by the stationarity condition which, at the tree level, reads 
\begin{equation}
v_\Delta = \frac{\nu}{\sqrt{2} } \frac{v_\phi^2}{M_{\Delta }^2} +\Or\!\left( \nu\frac{ v_ \phi^4}{M_\Delta^4}\right),
\label{HTM-TreeStatCondDeltaSolution}
\end{equation}
assuming, as usual, $M_\Delta^2 \gg v_\phi^2$. 
Subsequently, also $v_{\Delta}$ is real and, hence, there is no CP violation in the scalar sector.
This observation admits for a convenient splitting of the real and imaginary parts of $\phi^{0}$ and $\Delta^{0}$ into their scalar and pseudo-scalar components 
\begin{equation}
\phi^0=\frac{1}{\sqrt{2}}\left( \phi_S + i \phi_P\right),
\qquad
\Delta^0=\frac{1}{\sqrt{2}}\left( \Delta_S + i \Delta_P\right),
\label{eq:HTM:VEVs}
\end{equation}
which do not mix with each other through the interactions in~(\ref{HTMpot}).   

The tree-level physical spectrum of the model is given in Table~\ref{Table:HTMspectrum}. It consists of one Higgs boson $H$ with mass $\mathfrak{m}_{H} \sim v_\phi$, while all the other physical scalars have masses of the order of $M_\Delta$ and, thus, should  decouple in the $M_\Delta \rightarrow  \infty$ limit.
In what follows we shall argue that this picture remains intact even at the quantum level. 

\fulltable{\label{Table:HTMspectrum} The HTM scalar masses and charges including Goldstone modes computed in the tree-level approximation (up to terms of higher orders in $v_\phi^2/M_\Delta^2$).}
\br
Eigenstate  & $\mathfrak{m}^2_\mathrm{tree}$ & $Q^{CP}$
\\ \ns \mr
$\ket{w^\pm}=
\left(1-\frac{\nu ^2 v_\phi^2}{2 M_{\Delta }^4}\right)\ket{\phi^\pm}+\frac{\nu  v_\phi}{M_{\Delta }^2}\ket{\Delta^\pm}
$&0&$\pm 1$
\\
$\ket{z}=
\left(1-\frac{\nu ^2 v_\phi^2}{M_{\Delta }^4}
\right)\ket{\phi_\mathrm{P}}+\frac{\sqrt{2} \nu  v_\phi}{M_{\Delta }^2}\ket{\Delta_\mathrm{P}}$
& 0 & $0^-$
\\
$\ket{H}\;\, =
\left(1-\frac{\nu ^2 v_\phi^2}{M_{\Delta }^4}\right)\ket{\phi_S-v_\phi} + \frac{\sqrt{2} \nu  v_\phi}{M_{\Delta }^2}\ket{\Delta_S-v_\Delta}$ 
&$\left(2 \lambda _1-\frac{2 \nu ^2}{M_{\Delta }^2}\right) v_\phi^2$ & $0^+$
\\
$\ket{H^{\pm\pm}}=\ket{\Delta^{\pm\pm}}\phantom{\bigg)}$
&$M_{\Delta }^2+\frac{1}{2} \lambda _4 v_\phi^2$ & $\pm 2$
\\
$\ket{H^\pm}=
\left(1-\frac{\nu ^2 v_\phi^2}{2 M_{\Delta }^4}\right)\ket{\Delta^\pm}-\frac{\nu  v_\phi}{M_{\Delta }^2}\ket{\phi^\pm}
\phantom{\bigg)}$ 
&$M_{\Delta }^2+\left(\frac{\nu ^2}{M_{\Delta }^2}+\frac{2\lambda _4\!+\!\lambda _5}{4} \right)v_\phi^2$
& $\pm 1$
\\
$\ket{A}=
\left(1-\frac{\nu ^2 v_\phi^2}{M_{\Delta }^4}
\right)\ket{\Delta_\mathrm{P}}-\frac{\sqrt{2} \nu  v_\phi}{M_{\Delta }^2}\ket{\phi_\mathrm{P}}
$
&$M_{\Delta }^2+\left(\frac{2 \nu ^2}{M_{\Delta }^2}+\frac{\lambda _4\!+\!\lambda _5}{2} \right)v_\phi^2 $
& $0^-$
\\
$\ket{H'}=
\left(1-\frac{\nu ^2 v_\phi^2}{M_{\Delta }^4}\right) \ket{\Delta_S-v_\Delta} - \frac{\sqrt{2} \nu  v_\phi}{M_{\Delta }^2} \ket{\phi_S-v_\phi}$
&$M_{\Delta }^2+\left(\frac{2 \nu ^2}{M_{\Delta }^2}+\frac{\lambda _4\!+\!\lambda _5}{2} \right)v_\phi^2 $
& $0^+$\\
\br
\endfulltable

Needless to say, the full analysis of the relevant one--loop effective potential is rather difficult due to the notorious matrix logs in (\ref{VS}) which, given the relatively large number of fields around, can not be re-written in  terms of simple functions. On the other hand, as long as we focus solely to the masses of the neutral CP-even Higgs bosons, it is sufficient to consider a ``shortened'' version of $V_{\rm eff}$ in which all arguments corresponding to the charged fields (i.e., fields that are not allowed to develop VEVs) are set to zero\footnote{This is nothing but saying that a partial derivative of a function $f(x_{1},..,x_{n})$ of $n$ variables in the $i$-th direction at a given point $X$ in its domain can be evaluated in such a way that all the other coordinates $x_{j}$ with $j\neq i$ are set to their specific values $x_{j}=X_{j}$ first and only then the derivative is taken.}:
\begin{equation}
V_{\rm eff}^{\rm short}(\phi_{S}, \Delta_{S})\equiv V_\mathrm{eff}\left(\phi_S,\Delta_S,
\mathrm{all\,charged\,fields}=0 \right)\,.
\label{eq:HTM-short}
\end{equation}
It is important to notice that a hypercharge phase transformation with $\alpha=2\pi$ (which is a good symmetry of $V_\mathrm{eff}$ in the Landau gauge), corresponding to
\begin{equation}
\phi_S \rightarrow -\phi_S, \qquad \Delta_S \rightarrow \Delta_S\,,
\end{equation}
is a good symmetry of $V_{\rm eff}^{\rm short}(\phi_{S}, \Delta_{S})$ and, hence, $\phi_{S}$ enters the latter only quadratically.
Thus, by virtue of the usual trick~(\ref{eq:MasterEquation2})  for $\phi_{S}$ one arrives at 
\begin{equation}
\label{eq:HTM:M2scalarQuantum}
\fl
\mathcal{M}^2_S \equiv 
\left\langle \begin{array}{cc} 
\frac{\partial^2 V_{\rm eff}}{(\partial\phi_S)^2} &
\frac{\partial^2 V_{\rm eff}}{\partial\phi_S \,\partial\Delta_S} \\
\frac{\partial^2 V_{\rm eff}}{\partial\phi_S \,\partial\Delta_S}  &
\frac{\partial^2 V_{\rm eff}}{(\partial\Delta_S)^2}
\end{array}\right\rangle
= \left(\!\begin{array}{cc}
4 v_\phi^2 \frac{\partial^2 \hat{V}}{\left(\partial\phi_S^2\right)^2}
&
2 v_\phi \frac{\partial^2 \hat{V}}{\partial(\phi_S^2)\,\partial \Delta_S}
\\
2 v_\phi \frac{\partial^2 \hat{V}}{\partial(\phi_S^2)\,\partial \Delta_S}
&
\frac{\partial^2 \hat{V}}{(\partial v_\Delta)^2}
\end{array}\right)
\vline_{ 
\begin{array}{l}
\phi_S^2\rightarrow v_\phi^2 \\ \Delta_S \rightarrow v_\Delta
\end{array}
},
\end{equation}
where $\hat{V}$ is the ``$\phi_{S}$-quadratic'' form of $V_\mathrm{eff}^{\rm short}$, i.e.,
\begin{equation}
\hat{V}(\phi_S^2,\Delta_S)=V_{\rm eff}^{\rm short}(\phi_{S}, \Delta_{S})\,.
\label{eq:HTM-Vhat}
\end{equation}
The relevant derivatives above can be evaluated readily; neglecting terms linear in the small VEV $v_{\Delta}$ one arrives at
\numparts
\begin{eqnarray}
\frac{\partial^2 \hat{V}}{\left(\partial\phi_S^2\right)^2}
&=\frac{\lambda_1}{2}+\frac{A}{32 \pi ^2}\,,
\\
\frac{\partial^2 \hat{V}}{\partial(\phi_S^2)\,\partial \Delta_S}
&=
-\frac{ \nu }{\sqrt{2}}\left(1+\frac{B}{32 \pi ^2}\right)\,,
\\
\frac{\partial^2 \hat{V}}{(\partial v_\Delta)^2}
&=
\left(1+\frac{C}{32 \pi ^2}\right) M_\Delta^2 +\frac{1}{2}  (\lambda_4+\lambda_5)v_\phi^2 \,,
\end{eqnarray}
\endnumparts 
where the tree level contributions are fully displayed and the loop corrections are parametrized in terms of dimensionless factors $A,B,C$ which all scale as $O(\log M_\Delta^2)$.

The hierarchical structure of the matrix~(\ref{eq:HTM:M2scalarQuantum}) admits for writting the leading contribution to its smaller eigenvalue corresponding to the SM Higgs mass in the ``seesaw form''
\begin{eqnarray}
\mathfrak{m}^2_{H} \approx 4 \left[
\frac{\partial^2 \hat{V}}{\left(\partial\phi_S^2\right)^2}
- \left(
\frac{\partial^2 \hat{V}}{\partial(\phi_S^2)\,\partial \Delta_S}
\right)^2
\left(
\frac{\partial^2  \hat{V}}{(\partial v_\Delta)^2}
\right)^{-1}
\right] v_\phi^2
\\ \qquad \approx
\left[
2 \lambda_1-\frac{2 \nu ^2}{M_\Delta^2}
+\frac{1}{16 \pi ^2}\left(2 A +(C-2 B)\frac{ \nu ^2}{ M_\Delta^2}\right)
 \right] v_\phi^2\,.
\end{eqnarray}
As anticipated, irrespective of whether the loop corrections are included or not, the physical Higgs mass still turns out to be proportional to $v_\phi$.

There is perhaps one more comment worth making here: Notice that, unlike in Sections~\ref{sec:AbelianModel} and~\ref{sec:superconductor}, the second scalar at play here ($\Delta_{S}$) does not enter $V_{\rm eff}^{\rm short}$ purely quadratically and, hence, its VEV does not factorize out in Eq.~(\ref{eq:HTM:M2scalarQuantum}).
This, in turn, means that the mass of the physical heavy neutral eigenstate $H'$ is  \emph{not} proportional to $v_\Delta$ but climbs much higher, $\mathfrak{m}^2_{H'}\approx M^2_\Delta$.

\section{Conclusions}
The naturalness of the stipulated large hierarchy between the electroweak scale and virtually any new scale of beyond-Standard-Model physics has been for a long time one of the guiding principles of the BSM model building.
To this end, it has been often argued that, e.g., the Higgs mass is prone to receiving large corrections proportional to such a new scale $\Lambda$; as a matter of fact many new concepts such as supersymmetry were invented for the  purpose of saving it from such a ``danger''. 

Remarkably, a significant part of the community remained rather skeptical to this simple line of reasoning as, if not for anything else, there are at least three different layers of the argument that require a way more careful account. 
First, it is difficult to accept that the mass of the Higgs particle should be strongly sensitive to a scale that, from the perspective of the SM as a renormalizable theory, can be interpretted as nothing more than its cut-off. Even such a  trivial observation should make it clear that the entire spectrum of the SM including the Higgs must be driven by the same (electroweak) scale which, technically, means that the Higgs mass must be proportional to the corresponding VEV~$v$. 
The second layer then concerns the fate of $v$ itself, namely, its internal perturbative structure as a compound of the scalar potential quadratic curvature and possible huge terms proportional to $\Lambda$.  Depending on the personal preference this may or may not be disturbing as the situation here closely resembles the notorious issues with the interpretation of the cancellation of infinities within the standard renormalization procedure. 
Third, there is the twist that these questions can be reiterated with every next order in the perturbative expansion which may, naively, lead to concerns about ``stability'' of the calculations made at the preceding levels.

In the current work we elaborate on the first two points above by complementing the previous analysis~\cite{fun} in several directions. We  exemplify the fact that the Higgs mass $m_{H}$ is always (i.e., to all orders in the loop expansion) proportinal to its VEV on an extended series of specific simplified models in which the role of heavy dynamics is played by all ``reasonable'' types of fields (barring gravitons and Rarita-Schwinger fields) with spins 0, $\frac{1}{2}$ and 1, i.e., a heavy scalar, a heavy Majorana fermion and a heavy vector, respectively.  To this end, we show that the dependence of  $m_{H}$ on the heavy scale follows the expectation based on dynamical arguments such as the decoupling theorem.  
In doing so we advocate a slightly unconventional approach in which the scalar potential is parametrized in terms of non-linear (typically, gauge-invariant) field-space coordinates providing a simple connection between the residual $\Lambda$-dependence of the Higgs mass and the asymptotic behaviour of the physical scattering amplitudes. On top of that, it also renders the above-mentioned puzzle of the internal structure of the VEV totally irrelevant as it is not required to be inspected at all.

\section*{Acknowledgements}
The authors acknowledge the support from the Grant Agency of the Czech Republic (GA\v{C}R), Project No. 17-04902S and from the Charles University Research Center UNCE/SCI/013. We are indebted to Timon Mede for illuminating discussions and to Renato Fonseca and Martin Zdr\'ahal for reading through the manuscript.

\appendix

\section{Amplitudes}
\label{App:Amplitudes}
The $n$--th derivative of a function of a square of a variable generally reads
\begin{equation}
\left(\frac{\partial}{\partial\phi}\right)^n V(\phi^2)
= 
\sum _{k=0}^{\left\lfloor \frac{n}{2}\right\rfloor } 
	\frac{ 2^{n-2 k}\, n!}{k! \,(n-2 k)!} 
	\,\phi ^{n-2 k} \, V^{(n-k)}\!\left(\phi ^2\right).
\label{derivatives}
\end{equation}
Taking $n=1,2$ Eq.~(\ref{derivatives}) reproduces (\ref{eq:MasterEquation1}) and (\ref{eq:MasterEquation2}), while for a few higher $n$ the above formula yields:
\begin{eqnarray}\fl
\Gamma^{(3)}(0)=   \label{gamma3}
	12 v V''\!\left(v^2\right)
	+8 v^3 V^{(3)}\!\left(v^2\right),\\ \fl
\Gamma^{(4)}(0)= \nonumber
	12 V''\!\left(v^2\right) 
	+48  v^2 V^{(3)}\!\left(v^2\right)
	+16  v^4 V^{(4)}\!\left(v^2\right),\\ \fl
\Gamma^{(5)}(0)=	\nonumber
	\qquad\qquad\quad 120 v V^{(3)}\!\left(v^2\right) 
	+160  v^3 V^{(4)}\!\left(v^2\right)
	+32 v^5 V^{(5)}\!\left(v^2\right),  \\ \fl
\Gamma^{(6)}(0)= 	\nonumber
	\qquad\qquad\quad 120 \, V^{(3)}\!\left(v^2\right)	
	+720 v^2 V^{(4)}\!\left(v^2\right) 
	+480 v^4 V^{(5)}\!\left(v^2\right) 
	+64 v^6 V^{(6)}\!\left(v^2\right).
\end{eqnarray}
Generally, a set of equations for $\Gamma^{(n)}$ with $n=3,\ldots,N$ involves all $V^{(i)}(v^2)$ for $i= 2,\ldots, N$ and one can express $V''$ in terms of $\Gamma$'s and $V^{(N)}$. This, in the $N\rightarrow \infty$ limit, leads eventually to (\ref{eq:InvertedAmplitudes}).

\section{Evaluation of the U(1) effective potential}
\label{App:U1Veff}
The gauge field mass matrix $M_{\rm G}$ in the Abelian Higgs model in Sect.~\ref{sec:AbelianModel} is one--dimensional and the evaluation of $V_\mathrm{G}$ is thus trivial. 
The calculation of $V_\mathrm{S}$ is facilitated by exploiting the symmetry of the effective potential. This admits to restrict the complete field space to the following configurations
\begin{equation}
(\phi, \chi_1, \chi_2) \rightarrow ( \phi, \chi, 0)
\end{equation}
and restore the full dependence by means of (\ref{eq:ChiQuadraticInvariant}) at the end.
The scalar field--dependent mass matrix then simplifies to a block--diagonal form
\begin{equation}
M_\mathrm{S}^2 = \left(\!\!\begin{array}{ccc}
m_\phi^2 + 3 \lambda\phi^2 + \kappa\chi^2
&
2\kappa\phi\chi & 0 \\
2\kappa\phi\chi &
m_\chi^2+\kappa \phi^2+3\rho \chi^2 & 0\\
0&0&m_\chi^2 + \kappa \phi^2 + \rho\chi^2
\end{array}\!\!\right).
\label{eq:AHM:MS}
\end{equation}
The trace of the matrix function in (\ref{VS}) can be performed by summing over its eigenvalues. In total, one arrives at
\begin{equation}\eqalign{
V_\mathrm{eff} = V_0 &+ \frac{1}{256\pi^2}
\sum_{\pm}\left(A\pm\sqrt{B}\right)^2\left(\log\!\left[\frac{A\pm\sqrt{B}}{2\,\mu^2}\right]-\frac{3}{2}\right)\\
&+\frac{1}{64\pi^2} \left(m_\chi^2 + \kappa \phi^2 + \rho\chi^2\right)^2\left(\log\!\left[\frac{m_\chi^2 + \kappa \phi^2 + \rho\chi^2}{\mu^2}\right]-\frac{3}{2}\right)\\
&+\frac{3}{64\pi^2}\,e^4\chi^4 \left(\log\!\left[\frac{e^2\chi^2}{\mu^2}\right]-\frac{5}{6}\right),
}
\label{eq:U1Veff}\end{equation}
where
\begin{eqnarray}
A=m_\phi^2+m_\chi^2+ (\kappa +3 \lambda )\phi^2+ (\kappa +3 \rho )\chi^2,
\\
B=16\kappa^2 \phi^2 \chi^2 + \left(\left(\kappa-3\lambda\right)\phi^2 - \left(\kappa-3\rho\right)\chi^2 - m_\phi^2 + m_\chi^2\right)^2.
\end{eqnarray}

\section{Small Higgs mass loop corrections}
\label{App:Vecors}
We exploit the generalized custodial symmetry of the effective potential and perform the calculation in the following configuration:
\begin{equation}
\left(\phi_1, \phi_2, \phi_3, \phi_4, \chi_1, \chi_2\right)\rightarrow 
\left(0,0,\phi,0,\chi,0\right)\,.
\end{equation}

\subsection{Scalar loops}
Due to block-diagonal structure of $M_{S}$ (which is $6\times 6$ here) the calculation of $V_{S}$ proceeds along similar lines as that in~\ref{App:U1Veff}.
 Its second derivatives with respect to the quadratic invariants exhibit polynomial dependence on $m^2_\phi$ and $m^2_\chi$ which can be eliminated by using the \emph{tree level} tadpole cancellation conditions 
\begin{equation}
\rho v_\chi^2 + \kappa v_\phi^2 +m_\chi^2 = 0, 
\qquad
\kappa \chi^2 + \lambda\phi^2 + m_\phi^2=0,
\end{equation}
as the error committed is of a higher order in perturbation theory. The resulting formulas read
\begin{equation}\eqalign{\fl
\left\langle\frac{\partial^2V_\mathrm{S}}{(\partial\phi^2)^2}\right\rangle
=
\frac{\kappa ^2}{32 \pi ^2} \log\!\left[\frac{\kappa  v_\phi^2\!+\!\rho  v_\chi^2+m_\chi^2}{\mu ^2}\right]
+\frac{3 \lambda ^2}{32 \pi ^2} \log\!\left[\frac{\lambda  v_\phi^2\!+\!\kappa  v_\chi^2+m_\phi^2}{\mu ^2}\right]
\\+
\frac{1}{32\pi^2} \left(9 \lambda ^2 \log\!\left[\frac{2 \lambda  v_\phi^2}{\mu ^2}\right]+\kappa ^2 \log\!\left[\frac{2 v_\chi^2 \left(\lambda  \rho -\kappa ^2\right)}{\lambda  \mu ^2}\right]\right) + O\!\left(\frac{v_\chi^2}{v_\phi^2}\log v_\phi^2\right),
}\end{equation}
\begin{equation}\eqalign{\fl
\left\langle\frac{\partial^2V_\mathrm{S}}{\partial\chi^2\,\partial \phi^2}\right\rangle
=
\frac{\kappa  \rho}{32 \pi ^2}  \log\!\left[\frac{\kappa  v_\phi^2\!+\!\rho  v_\chi^2+m_\chi^2}{\mu ^2}\right]
+\frac{3\kappa \lambda}{32 \pi ^2}   \log\!\left[\frac{\lambda  v_\phi^2\!+\!\kappa  v_\chi^2+m_\phi^2}{\mu ^2}\right]
\\+ \frac{\kappa}{32\pi^2}\Bigg(
2 \kappa-\frac{2 \kappa ^2}{\lambda } 
+\frac{2 \kappa ^2+4 \kappa  \lambda +3 \lambda ^2}{\lambda }
\log\!\left[\frac{2 \lambda  v_\phi^2}{\mu ^2}\right]
\\+\frac{3 \lambda  \rho -2 \kappa ^2 
}{\lambda }
\log \left[\frac{2 v_\chi^2 \left(\lambda  \rho -\kappa ^2\right)}{\lambda  \mu ^2}\right]
\Bigg) + O\!\left(\frac{v_\chi^2}{v_\phi^2} \log v_\phi^2\right),
}\end{equation}
and
\begin{equation}\eqalign{\fl
\left\langle\frac{\partial^2V_\mathrm{S}}{(\partial\chi^2)^2}\right\rangle
=
\frac{\rho ^2 }{32 \pi ^2}
\log\!\left[\frac{\kappa  v_\phi^2\!+\!\rho  v_\chi^2+m_\chi^2}{\mu ^2}\right]
+\frac{3 \kappa ^2}{32 \pi ^2} \log\!\left[\frac{\lambda  v_\phi^2\!+\!\kappa  v_\chi^2+m_\phi^2}{\mu ^2}\right]
\\+\frac{1}{32\pi^2}\Bigg(
\frac{ \kappa ^2 \left(\lambda^2+12\lambda \rho -4 \kappa ^2\right) }{\lambda ^2}
\log\!\left[\frac{2 \lambda  v_\phi^2}{\mu ^2}\right]
\\+\frac{ \left(2 \kappa ^2-3 \lambda  \rho \right)^2 }{\lambda ^2}
\log\!\left[\frac{2 v_\chi^2 \left(\lambda  \rho -\kappa ^2\right)}{\lambda  \mu ^2}\right]
\\
+\frac{4 \kappa ^2 \left(2 \kappa ^2+\kappa  \lambda -3 \lambda  \rho \right)}{\lambda ^2}
\Bigg) + O\!\left(\frac{v_\chi^2}{v_\phi^2} \log v_\phi^2\right).
}\end{equation}
Clearly, those contributions grow only logarithmically with $v_\phi \gg v_\chi$, as supposed in the derivation of (\ref{eq:superCond_H_mass}\textit{,b}).
Note also that the spurious IR divergences in the above formulas can be dealt with as in~\cite{fun}.

\subsection{Vector boson loops}

Denoting the $SU(2)$ gauge fields by $A^a_\mu$ and the Abelian one by $B_\mu$ the gauge field mass matrix in the $\{A^1_\mu, A^2_\mu, A^3_\mu, B_\mu\}$ basis reads
\begin{equation}
M_\mathrm{G}^2 (\phi,\chi)= 
\frac{1}{4}
\left(\!\begin{array}{cccc}
 g^2 \phi^2 & 0 & 0 & 0 \\
 0 &  g^2 \phi^2& 0 & 0 \\
 0 & 0 & g^2 \phi^2 & - g g' \phi^2 \\
 0 & 0 & - g g' \phi^2  & g'^2 \!\left(16 \chi^2 + \phi^2 \right) 
\end{array}\!\right),
\end{equation}
and the gauge field contribution (\ref{VG}) to $V_\mathrm{eff}$ can be  calculated readily. Its second derivatives are
\begin{equation}
\fl \left\langle\frac{\partial^2V_\mathrm{G}}{(\partial\phi^2)^2}\right\rangle \approx \frac{g^4}{256\pi^2}  \left( 3 \log\!\left[\frac{g^2v_\phi^2}{4\mu^2}\right] +2 \right)
+\frac{\left(g^2+{g'}^2 \right)^2 }{512 \pi ^2}
\left(3\log\!\left[\frac{ \left(g^2+{g'}^2 \right)v_\phi^2}{4 \mu ^2}\right]+2\right),
\end{equation}
\begin{equation} \fl
\left\langle\frac{\partial^2 V_{G}}{\partial\phi^2\,\partial\chi^2}\right\rangle
\approx
\frac{{g'}^4 }{32 \pi ^2}
\left(3\log\!\left[\frac{ \left(g^2+{g'}^2 \right)v_\phi^2}{4 \mu ^2}\right]+2\right)
,
\end{equation}
and
\begin{equation}\eqalign{\fl
\left\langle\frac{\partial^2V_{G}}{(\partial\chi^2)^2}\right\rangle \approx
\frac{{g'}^4}{\pi ^2 \left(g^2+{g'}^2 \right)^2}
\,\Bigg\lbrace
3 {g'}^2  \left(\frac{1}{2} {g'}^2 +g^2\right) 
\log\!\left[\frac{\left(g^2+{g'}^2 \right)v_\phi^2 }{4 \mu ^2}\right]
 \cr
+ g^4- {g'}^2 g^2 +  {g'}^4 
+\frac{3}{2} \,g^4  
\log\!\left[\frac{4{g'}^2 g^2 v_\chi^2 }{\left(g^2+{g'}^2 \right)\mu ^2}\right]
\Bigg\rbrace 
\,
}
\end{equation}
where the higher-order $O\!\left(v_\phi^{-2}{v_\chi^2\log v_\phi^2}\right)$ terms have been neglected.

\end{document}